\documentclass[copyright,creativecommons]{eptcs}
\providecommand{\event}{SYNT 2015} % Name of the event you are submitting to

\usepackage[utf8]{inputenc}

\usepackage{graphicx}
\usepackage{amssymb}
\usepackage{amsmath}

\usepackage[htt]{hyphenat}
\usepackage{url}
\usepackage{breakurl} % for latex compatibility at EPTCS

\usepackage{xspace}
\usepackage{color}

\usepackage{listings}
\definecolor{DarkGreen}{rgb}{0,0.3,0}
\definecolor{DarkBlue}{rgb}{0,0,0.7}
\lstdefinelanguage{SMV}[]{}{
  commentstyle=\color{DarkGreen}\itshape,
  keywordstyle=\color{blue}\bfseries,
  morekeywords=[1]{AUXVAR,VAR,INIT,LTLSPEC,TRANS,GUARANTEE,ASSUMPTION,TRUE,X,G,F,CTLSPEC,AG,EX,MODULE,VARENV,DEFINE,LTLSPECENV,next,new,boolean},
  morekeywords=[1]{occurs,between,and,GF,leads,to,After,have,at,most,two,until,Globally,S,responds,
  after}, morecomment=[l]{--}
}

\lstdefinelanguage{MAA}[]{}{
  morekeywords=[1]{state,initial},
  morecomment=[l]{//}
}

\newcommand{\myTitle}{Synthesizing a Lego Forklift Controller in
GR(1):\\A Case Study}

\title{\myTitle}

\author{Shahar Maoz \qquad Jan Oliver Ringert
\institute{School of Computer
Science\\
Tel Aviv University, Israel}
\email{http://www.cs.tau.ac.il}
}

\def\titlerunning{\myTitle} % c 
\def\authorrunning{S. Maoz \& J.O. Ringert}

\newcommand{\true}{{\small{\op{true}}}\xspace}

\newcommand{\op}[1]{\textbf{\texttt{#1}}\xspace}
\newcommand{\blu}[1]{{\color{blue}\textbf{#1}}}

\newcommand{\vone}{\textbf{V1.Delay}\xspace}
\newcommand{\vtwo}{\textbf{V2.Continuous}\xspace}

\begin{document}

\maketitle

\begin{abstract}
Reactive synthesis is an automated procedure to obtain a correct-by-construction
reactive system  from a given specification.
GR(1) is a well-known fragment of linear temporal logic (LTL) where
synthesis is possible using a polynomial symbolic algorithm.
% c
We conducted a case study to learn about the challenges 
that software engineers may face when using GR(1)
synthesis for the development of a reactive
robotic system.
% c
In the case study we developed two variants of a forklift controller,
deployed on a Lego robot. The case study employs LTL specification
patterns as an extension of the GR(1) specification language, an
examination of two specification variants for execution scheduling,
traceability from the synthesized controller to constraints in the
specification, and generated counter strategies to support
understanding reasons for unrealizability.
% c
We present the specifications we developed, our observations, and
challenges faced during the case study. 

\end{abstract}

\section{Motivation and Context}

Reactive synthesis is an automated procedure to obtain a correct-by-construction
reactive system  from a given specification, if one exists. The time
complexity for synthesis of a reactive system from a linear temporal
logic (LTL) formula is double exponential in the length of the
formula~\cite{PR89}. However, limited fragments of LTL together with symbolic
implementations exhibit more practical time complexities. One such
fragment is General Reactivity of rank 1 (GR(1)), where
synthesis is possible using a polynomial symbolic
algorithm~\cite{BJP+12,PitermanPS06}.

The availability of efficient synthesis algorithms, as in the case of
GR(1), and the guarantee of implementations being correct by
construction, motivate applications in software
engineering.  GR(1) synthesis has been recently used in various application domains, including
robotics~\cite{Kress-GazitFP09}, scenario-based
specifications~\cite{MaozS12}, aspect languages~\cite{MaozS11}, and
event-based behavior models~\cite{DIppolitoBPU13}, to name a few.

We conducted a case study to explore the benefits and current challenges
of using existing tools and implementations of GR(1) synthesis to
develop a reactive system. The research objectives of our case study
were to learn about the following questions:

\begin{itemize}
  \item What are challenges faced when using a GR(1) synthesis tool to
  synthesize a software controller for a robotic system?
  \item Is the use of LTL specification patterns helpful to formulate
  assumptions and guarantees?
  \item Is it easy to understand reasons for unrealizability?
  \item Do successfully synthesized controllers work as expected? If
  not, how can one understand why?
\end{itemize}

Our goal is to learn what reactive GR(1) synthesis offers and what it
lacks to be successfully applied by software engineers in a model-based
development process. On a wider scale, we want to understand what is
required to make a synthesis specification language and synthesis tools
available and accepted by reactive systems software engineers.

It is important to note that our case study concerns the initial
development of a specification and synthesized controller. This
case study was not about applying new methods and tricks for
specification rewriting to make synthesis faster or to optimize
the synthesized controller, e.g., as in case studies of the AMBA
AHB protocol~\cite{BloemJK14,GodhalCH13}.
When including excerpts of
the developed specifications in this paper we decided to not rewrite
LTL formulas in a more \emph{clever} way but to present them as written
during specification development.

The context of our case study is the synthesis of software for a robotic
system. Synthesis is limited to a single controller that interacts with
its environment. The specification is written and analyzed by a software
engineer and the synthesized controller is used without modification for
automatic, direct code generation and deployment to a robotic system.
The tools used in our case study are:

\begin{itemize}
  \item a symbolic GR(1) synthesis algorithm implementation
  from~\cite{BJP+12} using the JTLV framework~\cite{PnueliSZ10}
  including the synthesis of counter strategies for unrealizable
  specifications; 
  \item the AspectLTL~\cite{MaozS11} input language, with syntax similar
  to SMV, for specifying environment assumptions and system guarantees,
  with syntax highlighting and code completion;  
  \item an implementation for traceability between a
  synthesized controller and its temporal
  specification based on~\cite{MaozS13AOSD}; and 
  \item a catalog of LTL specification patterns~\cite{DAC99} and
  their GR(1) templates~\cite{MR15patterns} (catalog of
  GR(1) templates available from~\cite{patternCatalog}.
\end{itemize}

The GR(1) synthesis problem is to find a controller that realizes a
given specification over a set of environment variables and a set of
system variables. A GR(1) specification consists of:

\begin{itemize}
  \item constraints on
initial assignments (constraints without temporal operators),
  \item safety constraints
over current and next assignments (constraints of the form \op{\blu{G}}
\texttt{(exp)} where the expression \texttt{exp} is limited to past time
temporal operators and the \op{\blu{next}} operator), and
  \item liveness
constraints, more specifically justice constraints, that should hold
infinitely often (constraints of the form \op{\blu{G F}} \texttt{(exp)}
where the expression \texttt{exp} has no temporal operators).    
\end{itemize}
 
All constraints are either assumptions, i.e., obligations of the
environment, or guarantees, i.e., obligations of the system.
Intuitively, if all assumptions are satisfied by the environment a
synthesized controller satisfies all guarantees. If no such controller
exists, the specification is called unrealizable. A detailed
introduction to LTL, past time LTL, and GR(1) synthesis is available
from~\cite{BJP+12}.

In this case study we use an extension of the GR(1) synthesis problem
where assumptions and guarantees may also be specified using LTL
specification patterns~\cite{DAC99}.
The extension is described in~\cite{MR15patterns}.
% The synthesis algorithm then computes a controller that implements the
% following specification:
% $$(\theta^e \wedge \op{G}\rho^e \wedge \bigwedge_{0<i\leq j}
% \op{GF}J_i^e \wedge \bigwedge_{j < i \leq m} \psi^e_i) \rightarrow
% (\theta^s \wedge \op{G}\rho^s \wedge \bigwedge_{0<i\leq k}\op{GF}J_i^s
% \wedge \bigwedge_{k < i \leq n} \psi^s_i)$$

Apart from the above GR(1) synthesis implementations we have used the
MontiArcAutomaton language~\cite{RRW14} for modeling the components of
the robotic system and as a concrete syntax for the synthesized
controllers. We have used code generators from this intermediate
representation to Java for deployment to the Lego Mindstorms NXT
platform\footnote{LEGO Mindstorms website:
\url{http://www.lego.com/en-us/mindstorms}}.

\section{Case Design and Variant}

\begin{figure}
  \centering
  \includegraphics[width=\textwidth]{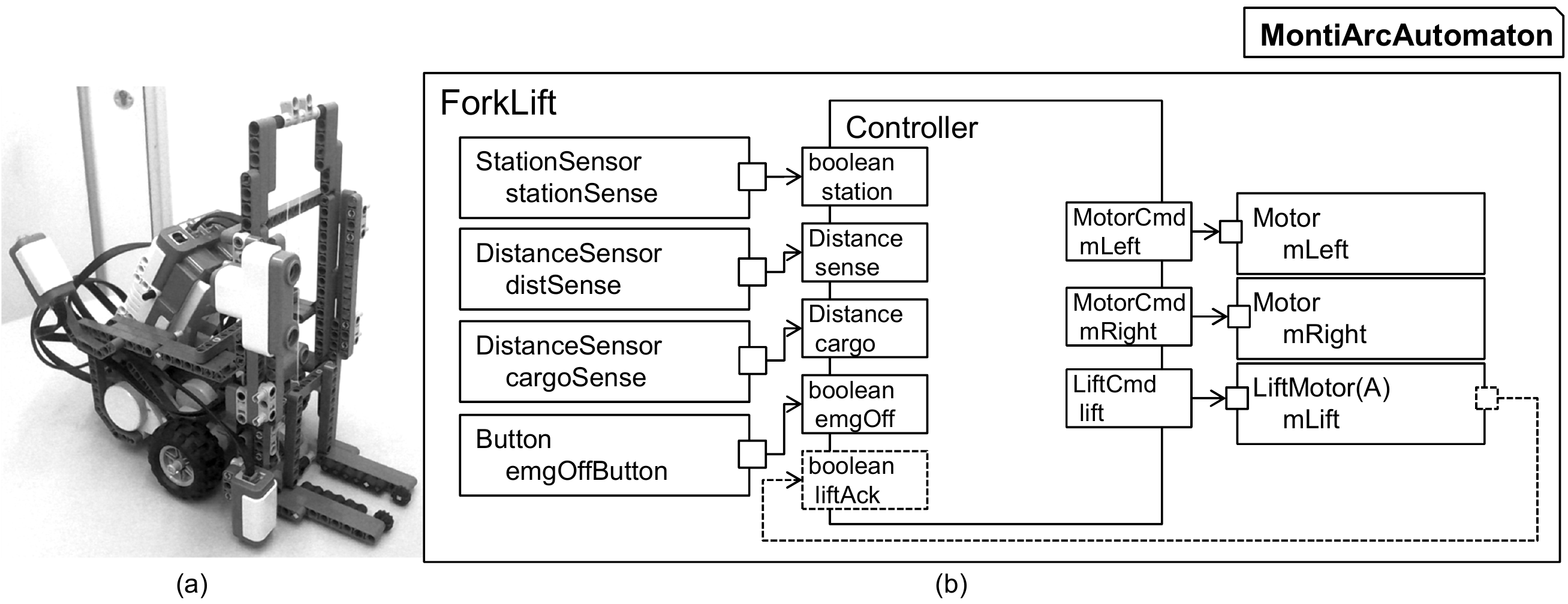}
  \caption{(a) The Lego forklift robot with four sensors and
  three actuators. (b) The logical software architecture of the robot
  with wrappers for sensors and actuators and the main component
  \texttt{Controller} to be synthesized (data types of input and
  output ports defined in Figure~\ref{fig:enums}).
  The dashed elements describe a second variant with feedback of the lift motor to acknowledge
  completion of lifting or dropping the fork}
  \label{fig:combined}
\end{figure}

The task of the case study was to develop a specification for a Lego forklift
robot shown on the left side of Figure~\ref{fig:combined}. The forklift is
an actual Lego robot\footnote{Robot based on these building
instructions: \url{http://www.nxtprograms.com/NXT2/forklift/steps.html}} we have
constructed and experimented with in our lab. A criterion for success of
synthesis was not only to obtain and inspect a synthesized controller
but also to deploy it to the real robot and see that its behavior
makes sense. 

The forklift shown in Figure~\ref{fig:combined} has a sensor to determine
whether it is at a station, two distance sensors to detect obstacles and
cargo, and an emergency button to stop it.
It has two motors to turn the left and right wheels and one motor to lift
the fork. The case definition consists of an initial set of informal
requirements for the behavior of the forklift:

\begin{enumerate}
  \item Do not run into obstacles.
  \item Only pick up or drop cargo at stations.
  \item Do not attempt to lift cargo if cargo is lifted.
  \item Always keep on delivering cargo.
  \item Never drop cargo at the station where it was picked up.
  \item Stop moving if emergency off switch is pressed.
\end{enumerate} 

Formalization and refinement of these requirements into guarantees and
the elicitation of suitable assumptions was part of the case study
execution. 

\begin{figure}
  \centering
  \includegraphics[width=.4\textwidth]{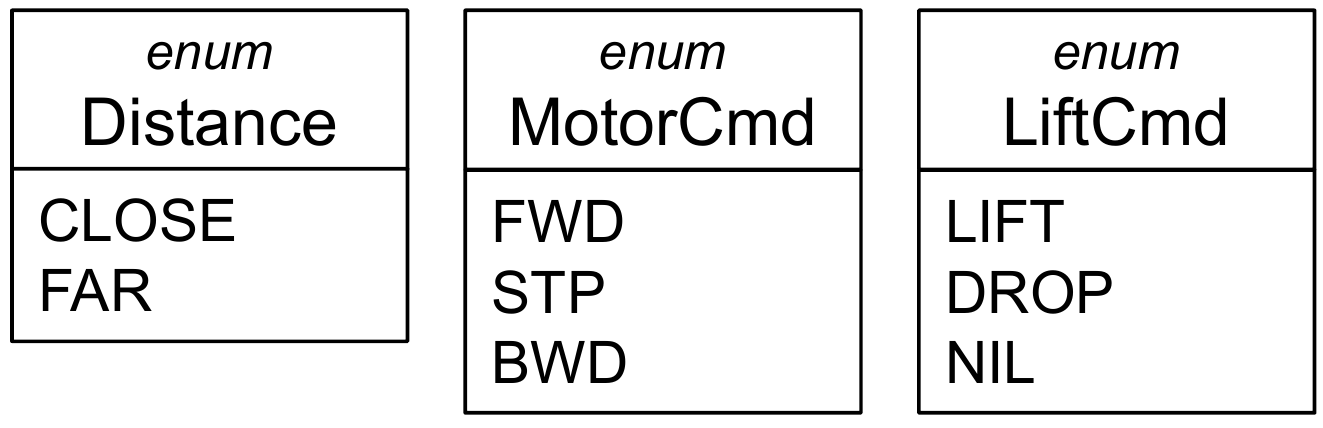}
  \caption{Data type definitions for ports of component
  \texttt{Controller} shown in Figure~\ref{fig:combined}}
  \label{fig:enums}
\end{figure}

Together with the above list of informal requirements the logical
software architecture of the forklift was defined before case study execution. It is depicted
as a component and connector model in 
Figure~\ref{fig:combined}~(b). The components on the left side are
hardware wrappers that read sensor values and publish them as messages on their
output ports. The output ports of the sensor components are connected to
input ports of component \texttt{Controller}. The output ports of
component \texttt{Controller} are connected to three components on the
right that receive commands and encapsulate access to the motors of the
forklift. The datatypes of input and output ports as well as their names
are written on the ports of component \texttt{Controller}. Datatypes
other than \texttt{boolean} are defined as enumerations in
Figure~\ref{fig:enums}.

The execution of the robot is performed in a cycle: read sensor data,
execute controller, perform actions.
When deploying the synthesized controller on the forklift robot the
output produced by transitions between states of the controller
manipulates the robots environment through its actuators.
We decided on two scheduling variants to integrate the synthesized
controller and the real world.

\begin{description}
\item[V1.Delayed] In the first variant the execution of transitions of
the synthesized controller is delayed to give the physical robot enough time
to execute tasks of the actuators. The delay has to be large enough to,
e.g., completely lift or drop the fork, complete a 90 degrees turn, or
back up from an obstacle. We set the delay to 2000ms.
\item[V2.Continuous] In the second variant the controller is
executed continuously without any delay.  This
variant uses a technique inspired by Raman et al.~\cite{RamanPK13} to
synthesize a reactive controller for continuous control.  The setting
requires the controller to be aware whether actions have completed or
not. In our case driving actions get feedback from
distance sensors but an additional feedback signal had to be added from
the lift motor to acknowledge completion of its actions (shown as a
dashed line in Figure~\ref{fig:combined}).
\end{description}
 
Depending on the scheduler and actuator implementations more variants
are possible, e.g., a variant where the scheduler pauses execution of
the synthesized controller while an actuator performs a
task (e.g., similarly by Kress-Gazit et al.~\cite{Kress-GazitFP07}).

%\jornote{maybe describe development steps carried out during
%synthesis}

\section{Resulting Specifications}

Development of the specifications started with the first variant
\vone in 7 versions with incremental addition of
features. When the first variant was almost complete the development of
variant \vtwo started based on the existing
specification. The final specifications for both variants are available
from~\cite{materials}.

The first and common part to both specifications is a schematic
translation of the input and output ports of component
\texttt{Controller} shown in Figure~\ref{fig:combined} to environment
variables (\texttt{VARENV}) and system variables (\texttt{VAR}) declared
in Listing~\ref{lst:interface}. The following excerpts of the two
specifications refer to variable names and short names defined in the
\texttt{DEFINE} block of Listing~\ref{lst:interface}. The only
difference between the variable declarations for \vone and \vtwo is the
environment variable \texttt{liftAck} in line~6 added only in variant \vtwo.

\begin{figure}
\lstset{language=SMV}
\begin{lstlisting}[label=lst:interface, caption={Environment variables
(\texttt{VARENV}), system variables (\texttt{VAR}), and definitions of
short names (\texttt{DEFINE}) of the controller to synthesize as an
implementation of component \texttt{Controller} from
Figure~\ref{fig:combined}~(b)}]
VARENV
  cargo : { CLEAR, BLOCKED };
  sense : { CLEAR, BLOCKED };
  station : boolean;
  emgOff : boolean;
  liftAck : boolean; -- only in V2.Continuous

VAR
  mLeft : { FWD, STOP, BWD };
  mRight : { FWD, STOP, BWD };
  lift : { LIFT, DROP, NIL };

DEFINE -- defines to ease reading specs
  backing := mLeft = BWD & mRight = BWD;
  stopping := mLeft = STOP & mRight = STOP;
  turning := mLeft = BWD & mRight = FWD | mLeft = FWD & mRight = BWD;
  forwarding := mLeft = FWD & mRight = FWD;
  dropping := lift = DROP;
  lifting := lift = LIFT;
\end{lstlisting}
\end{figure}

The two system variants are however very different. The main difference between
both systems is that the first variant of the forklift is scheduled in steps of
2000ms (it waits for 2 seconds after each read input, compute, write output
iteration), while the second variant does not use any delay. To illustrate this
difference: when executing the controller synthesized from \vone without the
delay the forklift runs over cargo it was expected to have lifted up; it may
initiate dropping cargo at a station but only completes dropping it after
leaving the station; during this process it detects the dropping cargo as an
obstacle and tries to drive clear from it.

\subsection{Specifications}

We give an overview of both specifications developed as part of the
case study.
We start with \vone and later discuss the changes in \vtwo compared to
\vone. 

\subsubsection*{\vone: Assumptions and Guarantees}

The specification document for \vone contains 7 assumptions, 12
guarantees, and 1 auxiliary variable. 

Of the 7 assumptions, 1 is a safety constraint, and 6 are LTL
specification patterns: 5 instances of the response pattern P26 and 1 instance of a
bounded existence pattern P15 (patterns numbered as appearing on the
website of~\cite{DAC99} and in our catalog of GR(1)
templates~\cite{MR15patterns,patternCatalog}). An example for a typical
response pattern is shown in Listing~\ref{lst:patternAsm}, ll.~1-3. The pattern expresses
the assumption that if the robot is backing or turning it will reach a state where both
distance sensors are clear unless it decides to go forward or stop, i.e., the
robot may \emph{escape} obstacles. A safety assumption about the
environment is that the reading of the sensor \texttt{station} will
not change when the forklift is stopping (see
Listing~\ref{lst:patternAsm}, ll.~5-6). Another assumption is
formulated using pattern P15 is shown in Listing~\ref{lst:patternAsm},
l.~8-9. It expresses that the robot will encounter at most two low
obstacles between stations.
% Both patterns can
% be translated to equivalent representations in GR(1) at the cost of
% adding additional variables to the statespace. Pattern P15 requires
% three auxiliary variables and every instance of pattern P26 requires
% one.

\begin{figure}
\lstset{language=SMV}
\begin{lstlisting}[label=lst:patternAsm, caption={Three assumptions of
variant \vone using LTL specification patterns~\cite{DAC99} with
equivalent representations in GR(1)~\cite{MR15patterns}}]
ASSUMPTION -- backing or turning clears sensors
  Globally (backing | turning) leads to 
               ((sense=CLEAR & cargo=CLEAR) | forwarding | stopping)); --P26
  
ASSUMPTION -- station does not change when stopping
  G (stopping -> station = next(station));

ASSUMPTION -- at most blocked twice between stations
  After (!atStation) have at most two (lowObstacle) until (atStation); --P15
\end{lstlisting}
\end{figure}

Of the 12 guarantees of \vone, 1 constrains the initial state, 8 are a
safety constraints, 1 is a justice constraint, and 2 are LTL specification
patterns (P09 and P20). One auxiliary variable is defined (we treat the
variable definition and assignments as a special case
although the assignments are currently implemented as four safety
guarantees). Our current implementation distinguishes manually added
auxiliary variables from other variables by the prefix
\texttt{spec\_} preceding the name of the variable.

\begin{figure}
\lstset{language=SMV}
\begin{lstlisting}[label=lst:patternGar, caption={Two guarantees of of
variant \vone using LTL specification patterns~\cite{DAC99} with
equivalent representations in GR(1)~\cite{MR15patterns}}]
GUARANTEE -- have to leave station to deliver 
  (!atStation) occurs between (lift=LIFT) and (lift=DROP);  --P09   

GUARANTEE -- don't drop cargo after leaving station until arriving 
  Globally (lift!=DROP) after (!atStation) until (atStation) --P20
\end{lstlisting}
\end{figure}

The specification uses pattern P09 as a guarantee that the forklift has
to leave the station between lifting cargo and delivering it
(Listing~\ref{lst:patternGar}, ll.~1-2).
% This pattern has a GR(1) representation that requires two auxiliary
% variables.
Another guarantee uses pattern P20 to express that after leaving a
station cargo cannot be dropped until the forklift reaches a station
(Listing~\ref{lst:patternGar}, ll.~4-5).

\begin{figure}
\lstset{language=SMV}
\begin{lstlisting}[label=lst:auxVar, caption={Definition of the
auxiliary variable \texttt{spec\_loaded}}]
VAR -- new auxiliary variable to "remember" when cargo is loaded
  spec_loaded : boolean;
GUARANTEE 
  !spec_loaded; -- initial value false
  G (lifting -> next (spec_loaded)); -- set loaded when lifting
  G (dropping -> ! next (spec_loaded)); -- unset loaded when dropping 
  G (lift = NIL -> next (spec_loaded) = spec_loaded); -- preserve value
    
\end{lstlisting}
\end{figure}

In Listing~\ref{lst:auxVar} a new auxiliary
variable \texttt{spec\_loaded} is introduced for the purpose of
simplifying the specification. This variable is used by the specifier to
keep track on whether the forklift has loaded cargo or not. The value of
the variable is determined by safety constraints. The
variable is used in the specification in both assumptions and
guarantees as shown in Listing~\ref{lst:usingLoaded}. The assumption in
lines~1-3 expresses that if the forklift goes forward and has cargo
loaded it will eventually find a station where it can deliver cargo
unless it stops or goes backward.

\begin{figure}
\lstset{language=SMV}
\begin{lstlisting}[label=lst:usingLoaded, caption={An assumption and a
guarantee of variant \vone using the auxiliary variable
\texttt{spec\_loaded} introduced in Listing~\ref{lst:patternGar}}]
ASSUMPTION -- find station to drop when forwarding (unless backing or stopping)
  Globally (forwarding & spec_loaded) leads to
                                 ((station & cargo=CLEAR) | backing | stopping));

ASSUMPTION -- find station and cargo when forwarding (unless backing or stopping)
  Globally (forwarding & !spec_loaded) leads to 
                               ((station & cargo=BLOCKED) | backing | stopping));

GUARANTEE -- do not lift again when cargo is loaded
  G (spec_loaded -> !lifting);
\end{lstlisting}
\end{figure}

Two additional guarantees in Listing~\ref{lst:goal} describe the
emergency stop feature of the forklift and the main justice constraint
to always eventually deliver cargo. The main justice constraint of the forklift is shown in
Listing~\ref{lst:goal}, ll.~4-5. It does not only contain the expected
\texttt{lift = DROP}, i.e., delivering cargo, but also the alternatives
\texttt{emgOff} and \texttt{lowObstacle}. These alternatives represent
environment actions that if occurring infinitely often liberate the
forklift from its obligation.

\begin{figure}
\lstset{language=SMV}
\begin{lstlisting}[label=lst:goal, caption={More guarantees of
variant \vone: the emergency stop feature and the main justice
constraint of the forklift}]
GUARANTEE -- emergency stop
  G (emgOff -> (stopping & lift=NIL));
   
GUARANTEE -- main goal to always eventually deliver cargo
  G F ((lift = DROP) | emgOff | lowObstacle);
\end{lstlisting}
\end{figure}

\subsubsection*{\vtwo: Assumptions and Guarantees}

We now describe the differences between the specification of the first
variant \vone and the second variant \vtwo of the forklift
specifications. The second variant applies a method inspired by Raman et
al.~\cite{RamanPK13} for continuous control. The main idea is to add a
new variable for every action which signals completion of the action. In
our case study this method is only necessary for the completion of
lifting and dropping cargo. The environment variable \texttt{liftAck} is
added (see Listing~\ref{lst:interface}) to signal completion of the
actions of component \texttt{LiftMotor} (see Figure~\ref{fig:combined}).

The specification document for variant \vtwo contains 9 assumptions, 14
guarantees, and 2 auxiliary variable definitions. None of the
assumptions or guarantees of \vone were removed, two
assumptions were added, and two guarantees were added.

\begin{figure}
\lstset{language=SMV}
\begin{lstlisting}[label=lst:newVarInV2, caption={Auxiliary
variable definitions in \vtwo: the assignments of
\texttt{spec\_loaded} were modified and variable
\texttt{spec\_waitingForLifting} was added}]
VAR   
  spec_loaded : boolean;
GUARANTEE -- updated assignments of var spec_loaded
  !spec_loaded;
  G (liftAck & !spec_loaded -> next(spec_loaded));
  G (liftAck & spec_loaded -> next(!spec_loaded));
  G (!liftAck ->  spec_loaded = next(spec_loaded));

VAR -- variable to keep track of waiting for lifting
  spec_waitingForLifting : boolean;
GUARANTEE
  !spec_waitingForLifting;
  G ((lift=LIFT | lift=DROP) -> next(spec_waitingForLifting));
  G (liftAck -> ((lift=LIFT | lift=DROP) | next(!spec_waitingForLifting)));
  G (!((lift=LIFT | lift=DROP) | liftAck) -> spec_waitingForLifting = next(spec_waitingForLifting));
\end{lstlisting}
\end{figure}

Of the 2 added assumptions, 1 is a safety constraint, and 1 is an
instance of pattern P26.
The auxiliary variable \texttt{spec\_waitingForLifting} is declared and
completely defined in Listing~\ref{lst:newVarInV2}, ll.~9-15 in the same
way as \texttt{spec\_loaded}. The variable is \true iff a lifting
command was issued and completion has not been acknowledged yet. The
assumption in Listing~\ref{lst:newAsmInV2}, ll.~1-2 expresses that an
acknowledgment \texttt{liftAck} will eventually follow every waiting. A
second assumption ensures that acknowledgments are only sent if
\texttt{spec\_waitingForLifting} is \true.
The assignment of variable \texttt{spec\_loaded} has been modified. The
assignment now depends on the previous value of the variable and the new
input \texttt{liftAck} (see Listing~\ref{lst:newVarInV2}, ll.~1-7).

\begin{figure}
\lstset{language=SMV}
\begin{lstlisting}[label=lst:newAsmInV2, caption={New assumptions added
in variant \vtwo about acknowledgments of executing the \texttt{LIFT}
and \texttt{DROP} commands}]
ASSUMPTION -- always eventually complete lifting
  Globally (spec_waitingForLifting) leads to (liftAck);
  
ASSUMPTION -- only acknowledge if waiting for it
  G (next(liftAck) -> (spec_waitingForLifting & !liftAck));
\end{lstlisting}
\end{figure}

Both added guarantees are safety constraints, as shown in
Listing~\ref{lst:newGarInV2}.  The first expresses that the
forklift stops when it waits for the completion of a lifting action (see
Listing~\ref{lst:newGarInV2}, ll.~1-2), and the second expresses that it
does not issue new lifting commands when waiting for completion (see
Listing~\ref{lst:newGarInV2}, ll.~4-5).

\begin{figure}
\lstset{language=SMV}
\begin{lstlisting}[label=lst:newGarInV2, caption={Added
guarantees of variant \vtwo}]
GUARANTEE -- while lift performs action stop 
  G (spec_waitingForLifting -> stopping);
 
GUARANTEE -- do not send another request when waiting for completion
  G (spec_waitingForLifting -> lift=NIL);
\end{lstlisting}
\end{figure}
\subsection{Synthesis Times and Controller Sizes}

For both variants a controller that implements the specification can be
synthesized in a few seconds using the Java-based BDD engine that comes
with JTLV~\cite{PnueliSZ10}. The sizes of both specification variants
are summarized in the upper half of Table~\ref{tbl:results}. We report
the assumption and guarantee constraints, the total number of variables
including auxiliary variables (the state space referred to as $N$ in the
time complexity $O(nmN^2)$ of GR(1) synthesis is 2 to the power of the
number of Boolean variables required to represent environment, system,
and auxiliary variables).
Environment and system variables amount to $2^{10}$ in \vone and to
$2^{11}$ in \vtwo.
For the specifications of variant \vone and \vtwo developed in this case
study 13 and 15 auxiliary variables were automatically (and implicitly)
added to support specification patterns or explicitly added as auxiliary
variables in the specification. The GR(1) synthesis problem after the
translation of patterns to GR(1), via the templates described
in~\cite{MR15patterns}, had 6 environment and 3 system justice goals in
variant \vone and 7 environment and 3 system justice goals in variant
\vtwo.

In the lower half of Table~\ref{tbl:results} we report the time it
took the synthesis algorithm to decide realizability and the additional
time consumed by the controller construction phase, in seconds. Times
are shown as reported by JTLV running on an ordinary laptop with Java 7,
Windows 7 64bit, 8GB RAM, and an Intel i7 CPU with 3.0 GHz. Synthesis
times for all intermediate versions during specification development
confirmed that synthesis is conveniently fast.

\begin{table}
\centering
% \begin{tabular}{l |r| r| r| r |r}
% Variant & Statespace & Justice Goals & Realizability & Constr. &
% \#States
% % & MontiArcAutomaton
% \\
% \hline
% \vone & $2^{23}$ & 6 env / 3 sys & 0.2sec & 1.8sec & 3412 
% %& 129 states, 1568 transitions
% \\
% \hline
% \vtwo & $2^{26}$ & 7 env / 3 sys & 0.7sec & 1.3sec & 2888 
% %& 115 states, 1408 transitions
% \\
% \end{tabular}
\begin{tabular}{l |r| r}
 & \vone & \vtwo \\
\hline
Assumptions &  1 safety, & 2 safety, 
\\
(patterns)&  5 times P26, P15 & 6 times P26, P15
\\
\hline
Guarantees &  1 initial, 8 safety, 1 justice, & 1
initial, 10 safety, 1 justice,
\\
(patterns)& P09, P20 & P09, P20\\
\hline
Boolean Variables & 4 environment, 6 system,  & 5 environment, 6 system,
\\
(auxiliary) & 1 manual, 12 pattern &  2 manual, 13 pattern
\\
\hline
Checking Realizability & 0.2 sec
& 0.7 sec 
\\
\hline
Controller Construction & 1.8 sec
& 1.3 sec 
\\
\hline
States of Controller & 3412 & 2888
\end{tabular}
\caption{For both variants we report the size of
the specification, times for checking realizability and controller
construction in seconds, and the size of a synthesized controller}
\label{tbl:results}
\end{table}

\subsection{Running Synthesized Controllers on the Forklift}

We used code generators implemented in our group for the NXJ LeJOS platform\footnote{Website of
LeJOS NXJ: \url{http://www.lejos.org/nxj.php}} to generate code and
directly deploy the software components shown in
Figure~\ref{fig:combined}~(b) to the LEGO NXT forklift shown in
Figure~\ref{fig:combined}~(a). The components \texttt{StationSensor}
(light value measured on the ground), \texttt{DistanceSensor}
(ultrasonic distance sensor), and \texttt{Button} (touch sensor) have
generic implementations in Java that wrap the LeJOS sensor API. The
components \texttt{Motor} and \texttt{LiftMotor} have generic
implementations that wrap the LeJOS motor API.

To execute the synthesized controller of variant \vone on the forklift
we had to adapt a set of platform specific parameters: number of 
degrees motors \texttt{mLeft} and \texttt{mRight} rotate backward and
forward, number of degrees the lifting motor rotates, distance to
obstacles detected by \texttt{distSense}, and distance to cargo detected by
\texttt{cargoSense}. In the second variant \vtwo the motors
\texttt{mLeft} and \texttt{mRight} do not rotate a fixed amount of
degrees but move continuously. 

The components on the robot are executed in execution cycles. Every
execution cycle starts with reading all sensor values, executes the
controller, and ends with executing all actuators. In variant \vone
we added a fixed delay of 2000ms after the execution of the actuators
to allow the forklift finishing all actions. In variant \vtwo a delay
was not added and the execution time of one cycle on the robot was
around 50ms.

We observed that variant \vone provided more reliable sensor readings
than \vtwo because it measured values after completing its actions in a
resting position.
For variant \vtwo erroneous sensor readings had serious effects. As one
example, when the station sensor falsely detected a station the forklift
stopped to drop cargo. A second reading when stopped did not report the
station and thus caused a violation of an environment assumption
(Listing~\ref{lst:patternAsm}, ll.~5-6), and so the forklift went into
an infinite loop of stopping. As another example, unstable readings of
the ultrasonic sensor led to detecting and not detecting obstacles at
the sensor's threshold level. This happened when the forklift was not at
a station and led to a violation of the assumption that at most two
obstacles occur (Listing~\ref{lst:patternAsm}, ll.~8-9). In this
case the forklift kept going forward in an infinite loop not stopping at
obstacles anymore.

Some works have addressed the challenge of synthesizing controllers
which are more robust to assumption violations, e.g., Ehlers and
Topcu~\cite{EhlersT14} suggested an approach where the synthesized
controller allows violations of safety assumptions up to some
constant number of times.

\section{Observations and Challenges}

We now report some of our observations and challenges faced during
specification development. 

\subsubsection*{O1: Differences between \vone and \vtwo}

The way that the synthesized controllers of variants \vone and \vtwo are
executed and interact with their environment is fundamentally different.
Thus, we found it surprising that their GR(1) specifications are still
very similar. Only two assumptions and two guarantees were
added. One reason for the similarity is that \vtwo
is based on \vone. It is still interesting that most existing
assumptions and guarantees also remain valid for the second variant,
although it is based on a continuous execution scheduling, without
delays. The main reason we did not have to adapt many assumptions seems
to be the use of the response patterns already in \vone (five instances)
instead of explicitly referring to immediate successor states.

\subsubsection*{O2: Manually adding auxiliary variables helpful} 

During the development of the two specifications we found it very
helpful to add auxiliary variables, as they assisted us in making
relevant states explicit. As an example, the auxiliary variable
\texttt{spec\_loaded} (Listing~\ref{lst:auxVar}) is defined to be \true
iff the forklift has loaded cargo. This information is derived
and not provided as a sensor input. The variable \texttt{spec\_loaded}
appears in two assumptions and two guarantees. The past LTL formula
% c
$$\texttt{\blu{PREV} (lift!=DROP \blu{SINCE} lift=LIFT)}$$
% c
could replace the auxiliary variable but we believe that it helps
readability of the formulas to make the property explicit with a new
name.

Technically, adding an auxiliary variable to the specification requires
that its value is determined by complete and deterministic safety
constraints. This mechanism is discussed by Bloem et al.~\cite{BJP+12}
for supporting past LTL in GR(1) synthesis.

\subsubsection*{O3: Support for patterns helpful}

LTL specification patterns~\cite{DAC99} allow one to express high-level
temporal patterns in a convenient way that are otherwise complicated
and error-prone to express in LTL. In the limited fragment of GR(1)
correctly expressing these behaviors becomes even more complicated due
to the limitation of available operators and no nesting. We provide GR(1) templates for
52 of the 55 LTL specification patterns~\cite{MR15patterns}.

During the case study we found that using patterns is very helpful for
expressing more complicated temporal properties. Moreover, using patterns gave us
better confidence that the specification matches our intention. 

A specifically useful pattern in assumptions was the response pattern
P26.
Instances of this pattern appeared as 5 out of 7 assumptions in \vone (6
out of 9 in \vtwo). This pattern seems to be well suited for describing
robotic systems where one should assume that some actions of actuators
eventually have an impact on sensor values.

It is important to note that the addition of patterns comes at a price. Most
patterns cannot be expressed without the addition of auxiliary variables. In
both variants of our case study more than half of the Boolean variables encoding
the statespace were auxiliary variables, implicitly added in the translation of
the patterns to the GR(1) form based on our templates. The resulting synthesis
problems of the case study are however still solved in a few seconds.

\subsubsection*{C1: Environment vs. real environment}

During the case study it turned out that it is difficult to use
assumptions to describe realistic environments. A very simple example
that appeared early during development of \vone is the assumption that
turning of the robot will make the distance sensor signal clear:
$$\texttt{\blu{G} (turning -> \blu{next}(sense=CLEAR));}$$
Our inspection of the synthesized controller concluded positively.
When deploying the controller to
the forklift it turned and stopped in a corner of the room. The assumption
was too strong for the forklift's real environment. In a corner, turning
90 degrees leaves the robot facing another wall. The
response pattern \texttt{\blu{Globally} (turning) \blu{leads to}
(sense=CLEAR)} would solve this particular problem but the
expressed assumption is still too strong and we observed another
undesired behavior: after being blocked the forklift turned once but
then stopped and waited for the obstacle to disappear. A corrected
version (also handling driving backwards) is shown in
Listing~\ref{lst:patternAsm}, ll.~1-3.

A different challenge appeared when adding a guarantee that the robot
should pick up cargo and not drop it between stations:
% c
\begin{multline*}
\texttt{\blu{Globally} (lift!=DROP \blu{SINCE} lift=LIFT)}\\
% b 
\texttt{\blu{after} (!atStation) \blu{until} (atStation);}
\end{multline*}
% c
The addition of this guarantee made the specification unrealizable. The
past time formula \texttt{lift!=DROP \blu{SINCE} lift=LIFT} embedded in the
specification pattern requires that the forklift leaves a station only
when cargo has been picked up. We decided that it might be reasonable to
assume that at every station the forklift may pick up cargo. It turned
out too complicated for us to formulate this assumption in GR(1). A
possible assumption might require encoding the area of each station and
valid navigation of the forklift to allow it to explore the station for
cargo. Finally, we weakened the above guarantee to the one in
Listing~\ref{lst:patternGar}, ll.~4-5. The modified version allows the
forklift to leave stations without cargo.

To summarize, in the case study we faced both the challenge of
formulating assumptions that are too strong for the real environment of
the forklift and the challenge of not being able to formulate reasonable
assumptions due to difficulties in expressing them. Both cases are not
easy to address. Specifically the first can lead to successful synthesis
of a controller that fails in a real environment.

\subsubsection*{C2: Undesired realizable case: system forces environment
to violate assumptions}

When the system can force the environment to violate its assumptions the
controller often does not act as the engineer would have expected it to
act. Consider the following two assumptions from an early version of
\vone. The first assumption is that if the forklift does not move the
station sensor value will not change. The second assumption is that the
forklift will eventually leave a station if it moves forward.
\begin{eqnarray}
\texttt{\blu{G} (stopping -> station = \blu{next}(station));}\nonumber\\
\texttt{\blu{Globally} (forwarding) \blu{leads to}
(!station);\nonumber}
\end{eqnarray}

We synthesized a controller and in some cases the forklift running the
controller stopped and did not continue to move. To understand this
behavior in the example above we enabled our synthesis tool to annotate
every transition with one of three reasons for it to be included in the
controller. The possible reasons are~\cite{BJP+12}: satisfying a justice
constraint, working towards satisfying a justice constraint, or
preventing a justice constraint of the environment.
The annotation of each transition of the controller includes the justice
constraint from the specification (see~\cite{MaozS13AOSD} for more
details about traceability).
Using this traceability information, which links transitions in the
controller to elements of the specification, we learned the reasons for the
synthesized strategy of the controller.
 We found out that in the example
above, the forklift goes forward on a station. If it is still on the
station in the next step it stops forever and thus prevents the
environment from satisfying the justice constraint of the response
pattern.

In this simple example with a controller of 80 states we were able to
find an explanation we could trace to the specification and modify the
response pattern as shown in Listing~\ref{lst:usingLoaded}, ll.~5-7.
It would have been helpful to automatically find cases where the system
forces the environment to violate assumptions for an engineer to decide
whether this behavior is desired or not. Towards the end of our
specification development process synthesized controllers simply became
too large for manual
 inspection.

Klein and Pnueli~\cite{KleinP10} defined environments where the system
cannot force a violation as \emph{well-separated}. They suggest a
modified GR(1) game to check whether an environment is well-separated. 
This may be a direction towards addressing this challenge.

\subsubsection*{C3: Unrealizable case}

In many cases adding a new feature expressed as a set of assumptions and
guarantees led to an unrealizable specification. In some cases we were
able to find mistakes quickly in the added assumptions and guarantees.
Many times we initially forgot to add alternatives in safety constraints
leading to their unsatisfiability by the system. In more complicated
cases we asked our tool to synthesize a counter strategy that represents
an environment forcing all system strategies to lose. We (interactively,
as described in~\cite{MaozS13AOSD,MaozS13}) executed moves of our
intended system and learned where our strategy loses against the
environment. This often led to better understanding of the reasons for
unrealizability.

In some cases it was however very difficult to understand the reason for
unrealizability, trace it back to the specification, and fix it.  As an
example, consider the introduction of the new environment variable
\texttt{liftAck} and the auxiliary variable
\texttt{spec\_waitingForLifting} (see Listing~\ref{lst:newVarInV2}). We
added the assumption that lifting can only be acknowledged if the
controller is expecting it:
% c
$$\texttt{\blu{G} (\blu{next}(liftAck) -> spec\_waitingForLifting);}$$

The specification of variant \vtwo with the above assumption is
unrealizable. A synthesized counter strategy has 3735 states. The Java
code generated by our synthesis tool for interactive exploration
of the counter strategy failed to compile due to its size. Printing the
counter strategy including information on successors disabled by safety
properties ran out of memory. It was not easily possible to reduce the
synthesis problem by removing irrelevant parts from the specification.
Lifting of cargo is related to movement of the motors and the cargo
sensor. The movement of motors is related to the distance sensor.

We manually executed the counter strategy by inspecting the generated
text output. Many steps were repetitive (long chains of apparently
similar states) as we were working towards forcing the environment to
present a station with cargo, and after lifting forcing it to
acknowledge lifting.
Right after acknowledging lifting the environment acknowledged lifting
again. The double acknowledgment set and immediately afterwards
disabled the variable \texttt{spec\_loaded} (see
Listing~\ref{lst:newVarInV2}, ll.~5-6). The double acknowledgment was
possible because \texttt{liftAck} disables
\texttt{spec\_waitingForLifting} only in the next step
(see Listing~\ref{lst:newVarInV2}, ll.~14). We adapted the
above assumption as shown in Listing~\ref{lst:newAsmInV2}, l.~5.

While counter strategies help understanding reasons of unrealizability
their handling by our tools turned out to be insufficient for larger
specifications.  In the future we plan to examine how recent work by
others, e.g.,~\cite{AlurMT13,KonighoferHB13}, may help in addressing the
unrealizability challenge.

\section{Conclusion}

We have presented a case study of the development of a software
controller for a forklift robot using GR(1) synthesis tools. Rather than
examining how to write most elegant and efficient specifications we
focused on challenges for software engineers in the process of
specification development.
We showed the specifications of two variants of the controller. On the
one hand, our observations are that extensions of the specification
language with auxiliary variables and higher-level specification
patterns support writing specifications with better confidence. On the
other hand, with growing specification size, understanding reasons for
synthesized behavior and for unrealizability turned out to be major
challenges.

This case study is part of our larger project on bridging the gap
between the theory and algorithms of reactive synthesis on the one hand
and software engineering practice on the other. In many aspects it
demonstrates the different challenges awaiting us.

\vspace{1em}
\noindent\textbf{Acknowledgments} The authors thank the anonymous reviewers of
the SYNT 2015 workshop for their helpful comments.
Jan O. Ringert acknowledges support from a postdoctoral Minerva Fellowship,
funded by the German Federal Ministry for Education and Research.
This project has received funding from the European Research Council (ERC) under
the European Union's Horizon 2020 research and innovation programme (grant
agreement No 638049, SYNTECH).

\bibliographystyle{eptcs}
\bibliography{doc}

\end{document}